\documentclass[twocolumn,floats,prl,superscriptaddress,showpacs,aps]{revtex4-1}
\usepackage{graphicx}%
\usepackage{dcolumn}%
\usepackage{bm}%
\begin{document}

\preprint{APS/123-QED}

\title{Electric-field-driven phase transition in vanadium dioxide}

\author{B. Wu}
\affiliation{Laboratoire de Physique et d'Etude des Mat\'eriaux, UMR 8213, ESPCI-ParisTech-CNRS-UPMC, 10 rue Vauquelin, 75231 Paris, France}
\affiliation{National Laboratory for Superconductivity, Beijing National Laboratory for Condensed Matter Physics and Institute of Physics, Chinese Academy of Sciences, Beijing 100190, People's Republic of China}
\author{A. Zimmers}
\email{Alexandre.Zimmers@espci.fr} \affiliation{Laboratoire de Physique et d'Etude des Mat\'eriaux, UMR 8213, ESPCI-ParisTech-CNRS-UPMC, 10 rue Vauquelin, 75231 Paris, France}
\author{H. Aubin}
\email{Herve.Aubin@espci.fr} \affiliation{Laboratoire de Physique et d'Etude des Mat\'eriaux, UMR 8213, ESPCI-ParisTech-CNRS-UPMC, 10 rue Vauquelin, 75231 Paris, France}
\author{R. Ghosh, Y. Liu and R. Lopez}
\affiliation{Department of Physics and Astronomy, University of North Carolina at Chapel Hill, Chapel Hill, North Carolina 27599, USA}

\date{\today}

\begin{abstract}
We report on local probe measurements of current-voltage and electrostatic force-voltage characteristics of electric field induced insulator to metal transition in VO$_2$ thin film. In conducting AFM mode, switching from the insulating to metallic state occurs for electric field threshold $\mathcal{E} \sim 6.5\times10^7 Vm^{-1}$ at $300~K$. Upon lifting the tip above the sample surface, we find that the transition can also be observed through a change in electrostatic force and in tunneling current. In this non-contact regime,  the transition is characterized by random telegraphic noise. These results show that electric field alone is sufficient to induce the transition, however, the electronic current provides a positive feedback effect that amplifies the phenomena.
\end{abstract}


\maketitle

Understanding the non-equilibrium transport properties of strongly correlated systems such as Mott insulators\cite{Mott1968} is a fundamental issue of condensed matter physics. This out of equilibrium many body problem presents a formidable challenge to theory\cite{Oka2005}; furthermore, the non-linear current voltage characteristics I(V) observed in correlated materials are of interest for the development of new electronic components; as exemplified recently with the discovery of the memristors\cite{Yang2008}.

Remarkable non-equilibrium transport properties due to electric field induced switching from insulating to a conducting regime have been observed in manganites\cite{Asamitsu1997}, vanadium dioxide (VO$_2$)\cite{Kim2005,Ko2008,Ruzmetov2009,Kim2010} and magnetite thin films or nanoparticles\cite{Lee2008}.

The material VO$_2$ is particularly interesting as it undergoes a first-order thermal induced Insulator to Metal Transition (IMT) at $340~K$\cite{Morin1959} where the resistivity changes by several orders of magnitude. The insulating state can be viewed as a solid of singlet vanadium atoms dimers in the Heitler-London limit, implying electron correlations\cite{Biermann2005}. The electrical field induced breakdown of this insulating state led to large current jumps (switching) in the I(V) curves\cite{Kim2005,Ko2008}.

The most important issue about this non-equilibrium transition is to understand the role of the electronic current. In non-equilibrium transport models based on electric field assisted tunneling such as in Poole-Frenkel or Zener mechanisms, only the electric field matters and the electronic current plays no role. In these models, the amplitude of the current can only increase exponentially fast with applied voltage. The switching phenomena -- which led to a faster than exponential increase of the current -- is puzzling as it implies either a many-body effect, such as a phase transition between an insulating and a metallic state, or some other non-linear effects due to avalanche breakdown or Joule heating.
While some experiments suggest that Joule heating could be responsible for the transition\cite{Kim2010}, others suggest this is negligeable\cite{Gopalakrishnan2009}.

To answer that question, we used an Atomic Force Microscope (AFM) to perform local measurements on a VO$_2$ film in the conducting AFM mode and in the Electrostatic Force Microscopy (EFM) mode. In the conducting AFM mode, the tip is in close contact with the sample and switching in the I(V) curve is observed. When the tip is retracted, we find that switching events can also be observed in the electrostatic force - voltage F(V) characteristics and in the tunneling current-voltage characteristics. In these last two modes, where no large current flows, the data present Random Telegraphic Noise (RTN), indicating that the system is bi-stable, jumping randomly from the insulating to the conducting state when an electrical field is applied. This shows that electric-field induced breakdown of a correlated insulating state is possible without any electronic current. However, the observation of RTN in those data, not observed when a large electronic current flows across the sample, suggests that the current plays a role in the transition through a positive feedback effect.

\begin{figure}[h!]
\begin{center}
\includegraphics[width=8cm,keepaspectratio]{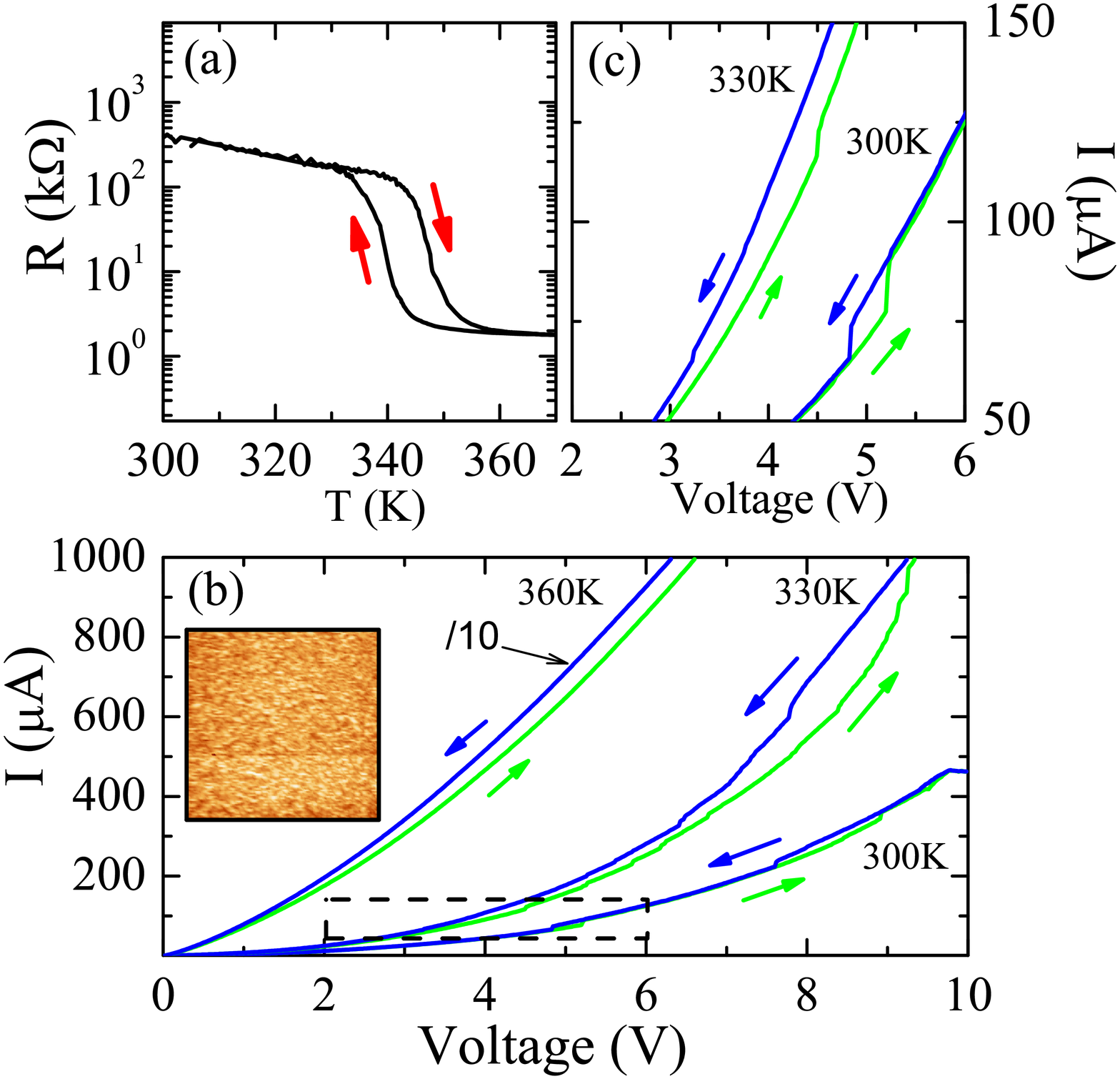}
\caption{\label{Fig1} (color online) Panel (a) Temperature dependence of the in-plane resistance of the VO$_2$ thin film deposed on doped silicon; Panel (b) I(V) curves measured at $300~K$, $330~K$, and I(V)/10 measured at $360~K$. Insert, AFM topography of the sample. Panel (c) Zoom on the current jumps, indicated as a dashed box on panel (b).}
\end{center}
\end{figure}

A $80~nm$ VO$_2$ thick film was deposited on a Si doped substrate by pulsed laser ablation \cite{Lopez2004}. AFM microscopy gives RMS surface roughness about $5~nm$ as shown in the insert of Fig.~\ref{Fig1}b. Figure~\ref{Fig1}a shows the sample resistance as the temperature is swept across the IMT. Local probe measurements are carried out with a home-built variable temperature AFM installed in Ultra High Vacuum. The force detection is based on measurement of frequency shift of a tuning fork\cite{Giessibl2003}. Typical amplitude oscillation of the tuning fork is about $5~nm$. The Pt/Ir tips glued onto the tuning fork were prepared by electrical etching or by mechanical cutting.

Figure~\ref{Fig1} shows typical $I(V)$ curves measured with the tip in contact with the surface, at different temperatures. At $300~K$, the first current jump is observed at voltage bias of $5.2~V$, which corresponds to an electric field threshold $\mathcal{E}=6.5\times10^7~Vm^{-1}$. As the temperature is increased, this threshold is shifted to lower voltage; at $T=330~K$, the first switching voltage is $4.5~V$, i.e. $\mathcal{E}=5.6\times10^7~Vm^{-1}$. These electric field thresholds are in good agreement with previous local probe measurements\cite{Ruzmetov2009,Kim2010} of switching current in VO$_2$. In contrast to measurements performed with macroscopic electrodes, our data can present several switches per I(V) curve. Probably, each additional current jump in the I(V) curve corresponds to the spreading of the transition area below the tip.

For temperature $T=360~K$, no current jumps are observed as expected for the sample in the metallic state.

\begin{figure}[h!]
\begin{center}
\includegraphics[width=7cm,keepaspectratio]{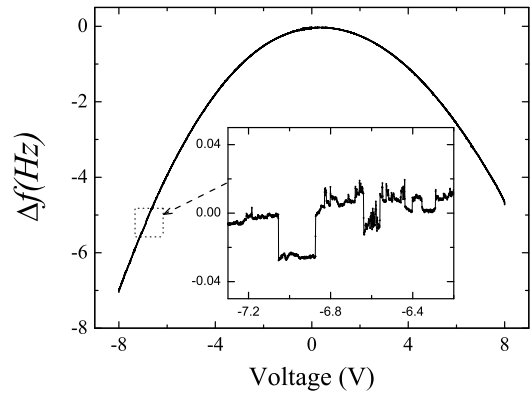}
\caption{Panel (a): Tuning fork frequency shift $\Delta$$f$ vs. tip-sample voltage difference $V$ for the tip located at 35nm above the surface. It shows the characteristic quadratic dependence of the electrostatic force with voltage. Small steps appear at high voltage, $\sim 7~V$, as shown in the inset with the parabolic background removed. They are the consequence of electric field induced IMT.\label{Fig2}}
\end{center}
\end{figure}

With this setup, one can now lift the tip and perform electrostatic force measurements. If the tip is far enough (more then 25nm), no current can be detected; however, the electrostatic force causes a shift of the tuning fork resonance frequency. Given that the electrostatic energy of the tip-sample capacitor C is $U_{el}=(1/2 C)\Delta \phi^2$, where $\Delta \phi=\phi_{tip}-\phi_{sample}+V$ is the electrochemical potential difference between the tip and the sample added to the applied voltage V, the electrostatic force is\cite{Colchero2001,Kurokawa1998}:

\begin{equation}\label{Eq2}
F=-\frac{1}{2}\frac{\partial C}{\partial z}\Delta \phi^2 \sim \ln{\frac{1}{z}}
\end{equation}

with the effective tip-sample distance $z=d+\lambda_D$, where $d$ is the tip-sample distance and $\lambda_D$ the Debye Screening Length (DSL).
Following Ref.~\cite{SeizoMoritaFranzJ.Giessibl2009}, the frequency shift $\Delta f$ is :

\begin{equation}\label{Eq1}
\Delta f=-\frac{dF}{dz}=\frac{1}{2}\frac{\partial^2 C}{\partial z^2}\Delta \phi^2 \sim-\frac{1}{z}
\end{equation}

The experimental quadratic behavior of the frequency shift vs. voltage is shown in Figure~\ref{Fig2}.

At tip-sample distance $d=35~nm$ and high voltage $\sim 7~V$, one observes small jumps in the electrostatic force $\Delta f$ vs. V curve. According to formula~\ref{Eq2}, these jumps could be interpreted either as changes in chemical potential or as changes in tip-sample capacitance.  As both the chemical potential and the DSL should be modified at the transition from the insulating to the conducting state, an electric field induced IMT provides a natural explanation for the observed jumps\footnote{Note that these jumps occur for either sign of the applied bias}. As a change in the DSL induces a change in the effective tip-sample distance $z$, its effect on the frequency shift can be estimated from $\delta\Delta f/\Delta f = -\delta z/z$ obtained from the derivation of Eq.\ref{Eq2}. The magnitude of the jumps are in the range $0.01-0.05~Hz$. As the full frequency shift due to electrostatic force is $\sim 5~Hz$, the relative change is about $0.2-1~\%$, which implies, with the tip-sample distance $d=35~nm$, a change of the DSL about $\simeq 0.07-0.35~nm$.

\begin{figure}[h!]
\begin{center}
\includegraphics[width=6.5cm,keepaspectratio]{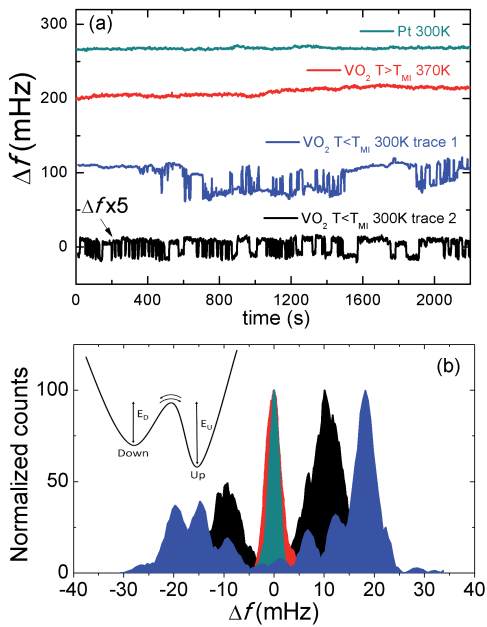}
\caption{(color online) Panel (a): Tuning fork frequency shift vs. time. From bottom to top, the first two curves are taken at two different places on the VO$_2$ sample at $300~K$; the third curve shows a time trace at $370~K$ in the metallic state, and finally, the upper curve is a time trace on a simple Pt thin film at $300~K$. Panel (b) Histograms of all time traces (using the same color code as panel (a)). Curves are normalized to set their maximum to 100 counts for each histogram. Insert panel (b): two-states fluctuator model discussed in text.\label{Fig3}}
\end{center}
\end{figure}

The DSL for VO$_2$ can be estimated from the carrier density n :
\begin{equation}\label{Eq3}
\lambda_D=\sqrt{\frac{\epsilon\epsilon_0 k_B T}{n e^2}}
\end{equation}
At the thermal-tuned IMT, the electron density changes by several orders of magnitude. In the $metallic$ state, Hall conductance measurements\cite{Ruzmetov2009a} give a carrier density about $n=10^{22}~cm^{-3}$, while capacitance measurements give $n=10^{19}~cm^{-3}$\cite{Yang2010}. Using the dielectric coefficient $\epsilon=30$\cite{Yang2010}, this led to a DSL in the range  $\lambda_D\simeq~0.07-2~nm$. In the $insulating$ state, capacitance measurements give a carrier density ranging from $n=5\times10^{15}~cm^{-3}$ to $n=5\times10^{18}~cm^{-3}$\cite{Yang2010}. This leads to a DSL in the range  $\lambda_D\simeq~3-90~nm$. With those values, we find that the change in DSL at the thermal tuned IMT is in the range $[1-90~nm]$. Thus, a change of DSL of one nanometer due to the electric field induced IMT is quite possible and provides a plausible explanation for the observed jumps in the frequency shift.

In contrast to the contact mode, we find that the electrostatic force detected transition is characterized by non-Gaussian noise in the frequency shift measured as function of time. Many time traces, for fixed tip position $35~nm$ above the surface, lasting about one hour, have been taken which all present RTN~\footnote{A drift, less than 1Hz/hour, was subtracted on each curve.} as shown by the bottom curve of Fig.~\ref{Fig3}a. A histogram of each time trace, Fig.~\ref{Fig3}b, shows that this noise is obviously non-Gaussian at 300K with two peaks corresponding to two stable states. We checked that the observed RTN could not arise from the instrument. First, RTN is never observed when measurements are performed on simple metal Pt surface, see Fig.~\ref{Fig3}a, and second,  RTN disappears when the temperature is increased at 370K above the IMT transition of VO$_2$. In those last two cases, we find instead that the noise has a Gaussian distribution, Fig.~\ref{Fig3}b.
RTN such as observed here is the characteristic signature of a two-states fluctuator. Two-states fluctuators can emerge from localized states of nanoscale devices or nanoparticles\cite{Cockins2009}, where fluctuations between the two states is due to trapping-detrapping of single charges. However, two-states fluctuators can also emerge from systems with many degrees of freedom. Typical examples are provided by Josephson junctions\cite{Leggett1987} or superconducting nanowires\cite{Sahu2009} where the system switches from the superconducting to the normal state when the superfluid current exceeds some critical value. Furthermore, RTN has also been observed in transport measurements of strongly correlated material such as underdoped cuprates\cite{Bonetti2004} and manganites\cite{Raquet2000}, interpreted as two-levels fluctuators emerging in systems with many degrees of freedom. In our system, the two-states fluctuator is built up from the two macroscopic states of VO$_2$ : one is insulating, one is conducting.

It has a double potential well structure \cite{Raquet2000,Hess2001}, as drawn schematically in the insert Fig.~\ref{Fig3}b. RTN is produced as thermal excitations drive the system randomly in one state or the other. In this model, the average time spent in one state $i$ is given by the Arrhenius law\cite{Kogan1996}: $\tau_i=\tau_0 \exp{\frac{E_i}{k_BT}}$ where $E_i$ is the energy of the state $i$ and $\tau_0$ is a microscopic constant which depends on the thermal coupling to the external bath. Using this formula, one finds the energy difference $E_U-E_D=k_B T \ln \frac{\tau_U}{\tau_D}=17~meV$, where the occupation times ${\tau_U=5.5~s}$ and ${\tau_D=2.9~s}$ are obtained\cite{Kogan1996} from counting the occupation times in the lowest time trace Fig.~\ref{Fig3}a.

Note that this energy level difference should depend on temperature and applied electric field. Furthermore, the detailed characteristics of the noise may change from one sample location to another one, as exemplified by the time trace 1 taken on another location of the sample, shown Fig.~\ref{Fig3}a. Interestingly, recent theoretical works\cite{Loh2010} have shown that electronic inhomogeneities such as stripes could lead to RTN, whose detailed  characteristics could reveal whether such noise is due to correlated fluctuations rather than independent switcher. However, such an analysis will require a more detailed spatial mapping of RTN than presented here.

\begin{figure}[h!]
\begin{center}
\includegraphics[width=8cm,keepaspectratio]{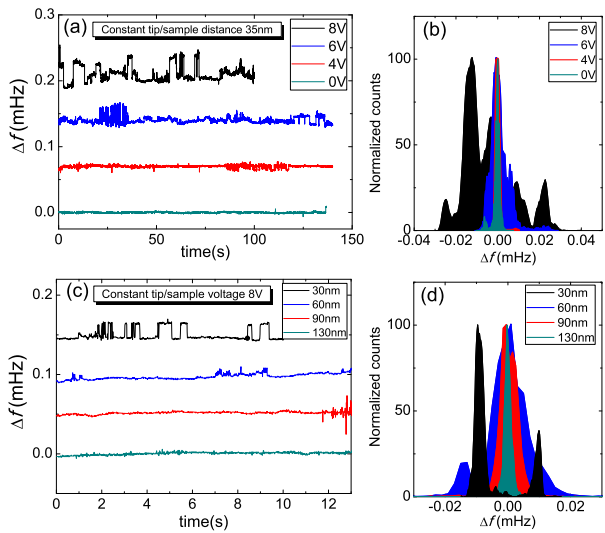}
\caption{(color online) EFM time traces for different electrical field values at 300K. Panel (a), the tip/sample distance is set to $35~nm$ and voltage is changed from 8V to 0V. Panel (c), the tip/sample voltage is set to $8~V$ and the tip/sample distance is varied from $30~nm$ to $130~nm$. Panels (b) and (d) show the histograms of each time traces presented in panels (a) and (c) respectively. Non-Gaussian noise is only found for tip/sample distance below 90nm and tip/sample voltage above 4V. \label{Fig4}}
\end{center}
\end{figure}

Figure~\ref{Fig4} shows that RTN only appears above some electric field threshold. For a tip-sample distance of $35 nm$, RTN disappears for voltage values lower than $\sim 4~V$. On the other hand, for a fixed voltage value $\sim8~V$, RTN disappears for tip-sample distance larger than $90~nm$. Using for the DSL the value estimated above, we find that the electric field threshold is of the order of $\mathcal{E}\simeq 10^8 Vm^{-1}$, which is comparable to the electric field threshold found for current switching in the contact mode.

\begin{figure}[h!]
\begin{center}
\includegraphics[width=6.5cm,keepaspectratio]{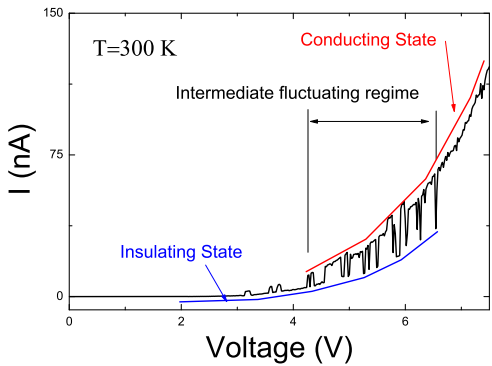}
\caption{(color online) I(V) curve in the tunnel regime.
\label{Fig5} }
\end{center}
\end{figure}

An intermediate situation between good tip-sample contact, shown Fig.~\ref{Fig1}, and purely electrostatic mode, shown Fig.~\ref{Fig3}, is the tunnel regime. In this regime, RTN is also observed in the current-voltage characteristics, shown Fig.~\ref{Fig5}. At low voltage, the sample is in the insulating state and at high voltage, the sample is in the conducting state. However, on a wide intermediate electric field range, the system switches randomly from the metallic to the insulating state, producing the observed RTN.

The observation of RTN in both the electrostatic force and tunnel current but not in the I(V) curves measured in the contact mode point toward an explanation of the role of the current in the transition. When the electronic current is allowed to flow, it induces some positive feedback -- possibly through Joule heating -- and drives the system deeper into the metallic state. On the other hand, if the current is not allowed to flow, as there is only a small free energy difference between the insulating and conducting states, the system keeps switching between the two states on a large electric field range.

To conclude, we have presented conducting AFM and EFM on VO$_2$ thin film. In the contact mode, current switching is seen in the I(V) curves. With no Joule heating (tunnel mode and EFM mode), the system does not have any positive feedback and RTN is observed in the data. The observation of these fast switching events in the absence of Joule heating indicates that many-body effects should be at the origin of this phenomena. Such a many-body phenomena could involve an electric field induced IMT or possibly some many-body versions of Zener tunneling as recently studied theoretically\cite{Oka2005}.

A.Z. and H.A acknowledge support from ANR grant 09-BLAN-0388-01.
R.G, Y.L. and R.L. acknowledge support from UNC-Chapel Hill Institute for the Environment Carolina Energy Fellows Program.
\bibliography{VO2paper}

\end{document}